\documentstyle[multicol,aps,prl,epsf]{revtex}
\begin{document}
\title{Breakdown of Scaling in the Nonequilibrium Critical Dynamics 
of the Two-Dimensional XY Model}
\author{A. J. Bray, A. J. Briant, and D. K. Jervis}
\address{Department of Physics and Astronomy, The University, Manchester,
M13 9PL, United Kingdom}

\date{\today}
\maketitle

\begin{abstract}
The approach to equilibrium, from a nonequilibrium initial state, in a 
system at its critical point is usually described by a scaling theory 
with a single growing length scale, $\xi(t) \sim t^{1/z}$, where $z$ is 
the dynamic exponent that governs the {\em equilibrium} dynamics. We show 
that, for the 2D XY model, the rate of approach to equilibrium depends 
on the initial condition. In particular, $\xi(t) \sim t^{1/2}$ if no free 
vortices are present in the initial state, while 
$\xi(t) \sim (t/\ln t)^{1/2}$ if free vortices are present.  
\end{abstract}

\begin{multicols}{2}
While the theory of equilibrium critical phenomena has been a mature subject 
for more than 20 years, nonequilibrium critical phenomena still pose some 
interesting challenges. The simplest scenario consists of a system 
evolving at its critical point from a nonequilibrium initial state in which 
the system was prepared at time $t=0$. Since the characteristic relaxation 
time is infinite at criticality, an infinite system will never reach 
equilibrium. Instead, the system evolves towards equilibrium through a 
nonequilibrium scaling state. Consider, for example, the equal-time 
pair correlation function, 
$C(r,t) = \langle \phi({\bf x},t)\,\phi({\bf x}+{\bf r},t)\rangle$, 
where $\phi$ is the order-parameter field.  
In the nonequilibrium scaling state it has the form 
\begin{equation}
C(r,t) = \frac{c}{r^{d-2+\eta}}\,f\left(\frac{r}{\xi(t)}\right),
\label{scaling}
\end{equation}
where $d$ is dimension of space, $\eta$ the usual critical exponent, 
and $c$ is a constant. The scaling form (\ref{scaling}) holds 
in the limit $r \gg a$, $\xi(t) \gg a$, with $r/\xi(t)$ arbitrary, where 
$a$ is a microscopic cut-off, e.g.\ a lattice spacing. The first factor 
in (\ref{scaling}) is the equilibrium correlation function: Requiring 
that this be recovered for $t = \infty$ forces $f(0)=1$. 

The physical interpretation of $\xi(t)$ is the length scale up to which 
critical correlations have been established at time $t$: 
$C(r,t) \sim c\,r^{-(d-2+\eta)}$, the equilibrium result holds, 
for $a \ll r \ll \xi(t)$. Dynamical scaling suggests 
\begin{equation}
\xi(t) \sim t^{1/z}
\label{xi}
\end{equation}
for large $t$, where $z$ is the usual dynamic exponent 
characterising temporal correlations {\em in equilibrium}. This 
result has been demonstrated in an expansion in $\epsilon = 4-d$ using 
standard field-theoretic renormalization group methods \cite{Janssen}. 
The importance of this result is that it shows that relaxation {\em to} 
equilibrium is governed by the same exponent as correlations {\em in} 
equilibrium. A second important result of reference \cite{Janssen} is 
that the relation $\xi(t) \sim t^{1/z}$ holds independently of the 
nonequilibrium initial state, which can affect the scaling function, 
$f(x)$, in (\ref{scaling}) but not the exponent $z$ (since this is a 
property of the equilibrium renormalization group fixed point). 

Two special cases illustrate the dependence of $f(x)$ on the initial 
conditions. For a disordered initial condition, the system will remain 
disordered on scales $r \gg \xi(t)$, so $f(x)$ will fall off rapidly for 
$x \gg 1$. For an initial condition with long-range order 
(i.e.\ non-zero initial magnetization), dynamical scaling predicts 
that the magnetization $M(t)$ will decay asymptotically as 
$t^{-\beta/\nu z} \sim \xi(t)^{-\beta/\nu}$. In this case $C(r,t)$ 
approaches $C(\infty,t) = M^2(t) \sim t^{-2\beta/\nu z} 
= t^{-(d-2+\eta)/z}$ using standard scaling laws. 
So in this case $f(x) \sim x^{d-2+\eta}$ for $x \to \infty$. 

The purpose of this Letter is to challenge this simple picture for the 
XY model in $d=2$ at (and below) the Kosterlitz-Thouless (KT) transition 
\cite{KT}. Specifically we argue that the growing length scale $\xi(t)$ 
satisfies (\ref{xi}), with $z=2$, if the initial state contains no free 
vortices, whereas $\xi(t) \sim (t/\ln t)^{1/2}$ if free vortices are 
present. We first present numerical simulation results supporting 
this scenario, and then provide the theoretical interpretation. 
The two types of initial state we shall consider are (a) completely 
ordered (no free vortices), and (b) completely random (free vortices 
present). 

The XY model consists of planar spins $\{\vec{S}_i\}$ at the sites of 
a square lattice of linear size $L$, with Hamiltonian 
$H = -\sum_{<i,j>}\vec{S}_i\cdot\vec{S}_j$, where the sum is over lattice 
links and we have taken the exchange interaction to have strength unity. 
We adopt conventional `heat-bath' dynamics in which a spin is moved to 
a trial configuration chosen at random on the unit circle, and the 
move accepted with probability $[1 + \exp(\Delta E/T)]^{-1}$, where 
$\Delta E$ is the energy change associated with the move, and $T$ 
is the temperature. The lattice is divided into two sublattices, and the 
sublattices updated alternately. One unit of time corresponds to an 
attempted move of every spin. 

A convenient quantity to study is the `time-dependent Binder cumulant' 
\cite{BHB,Zheng}, $g_L(t)$, defined by 
\begin{equation}
g_L(t) = 2 - \frac{\langle(\vec{M}^2)^2\rangle}{\langle\vec{M}^2\rangle^2},
\label{Binder}
\end{equation}
where $\vec{M}(t) = \sum_i \vec{S}_i(t)$ is the total magnetization at 
time $t$, and $\langle \ldots \rangle$ indicates an average over independent 
Monte Carlo runs ($10^4$ runs were used in practice). 
Because the powers of $\vec{M}$ in numerator and denominator 
are the same, $g_L(t)$ depends (at a critical point, and if dynamical 
scaling holds) only on the ratio $\xi(t)/L$, where $L$ is the (linear) 
size of the lattice:
\begin{equation}
g_L(t) = G\left(\frac{\xi(t)}{L}\right),
\label{g-scaling}
\end{equation}
provided, as always, that both $\xi(t)$ and $L$ are sufficiently large. 
This result provides the basis for a determination of $\xi(t)$ using 
finite-size scaling. If conventional dynamical scaling holds, the scaling 
{\em function} $G(x)$ may depend on the initial state, but the scaling 
{\em variable},  $\xi(t)/L = t^{1/z}/L$, will not. For an ordered initial 
state (all spins parallel) $G(0)=1$, while for a random initial condition 
(each spin chosen independently from a unit circle) $G(0)=0$ follows from 
the Gaussian distribution (central limit theorem) of $\vec{M}(0)$. 
For $t \to \infty$, $G(x)$ approaches, in both cases, the universal value 
$G(\infty$) characteristic of the critical point. For the KT phase there 
is actually a line of such fixed points, $T \le T_{KT}$, 
(and a corresponding set of values $G_T(\infty)$) but we will focus 
primarily on the KT point, $T_{KT}$, 
using the accepted value $T_{KT} = 0.90$ \cite{Tc}.   
  
Data for the ordered initial state are presented in Figure 1, for system 
sizes $L=12$, 16, 24, 32, and 48. The abscissa, $t/L^2$, corresponds to 
a scaling variable $t/L^z$ with $z=2$. This choice of $z$ is dictated 
by the spin-wave theory (i.e.\ no free vortices) that describes the 
large-scale properties of the KT phase everywhere along the fixed line 
$T \le T_{KT}$. The best collapse using all the data favors a 
slightly lower value, but the value $z=2$ clearly gives a good scaling 
collapse for larger $L$, i.e.\ $L \ge 24$ (note the expanded scale 
compared to Figures 2 and 3). Collapsing the data for pairs 
of $L$ gives effective exponents $z(L_1,L_2)$ given by $z(12,16)=1.75(5)$, 
$z(16,24)=1.83(3)$, $z(24,32)=1.96(2)$, $z(32,48)=2.00(2)$, consistent 
with a convergence to $z=2$ for $L \to \infty$. Recent simulations by Luo 
et al.\ \cite{Zheng2} give similar results: $z=1.96(4)$ for $T=0.90$ and 
an ordered initial state.  

\begin{figure}
\narrowtext
\centerline{\epsfxsize\columnwidth\epsfbox{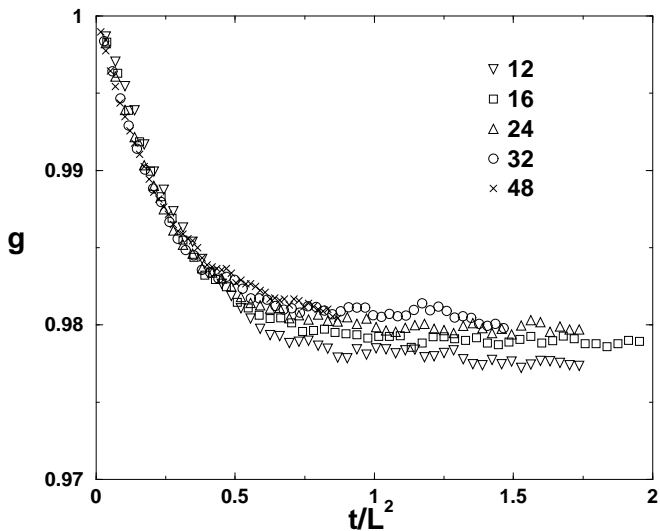}}
\caption{Scaling plot, with $z=2$, for the time-dependent Binder 
parameter, starting from an ordered initial condition, for system sizes 
$L=12,16,24,32,48$}. 
\label{order}
\end{figure}

The date for a random initial condition are presented in Figures 2 and 
3. In Figure 2, we attempt to collapse the data with a scaling variable 
$t/L^z$. The scaling collapse is very good, but a much higher value of 
the dynamical exponent, $z \simeq 2.35$, is required than for an ordered 
initial state. For a random initial condition Luo et al.\ found, by direct 
measurement of the time-dependence for a large lattice ($L=512$), the 
slightly smaller result $z=2.29(1)$ \cite{Zheng2}.   

\begin{figure}
\narrowtext
\centerline{\epsfxsize\columnwidth\epsfbox{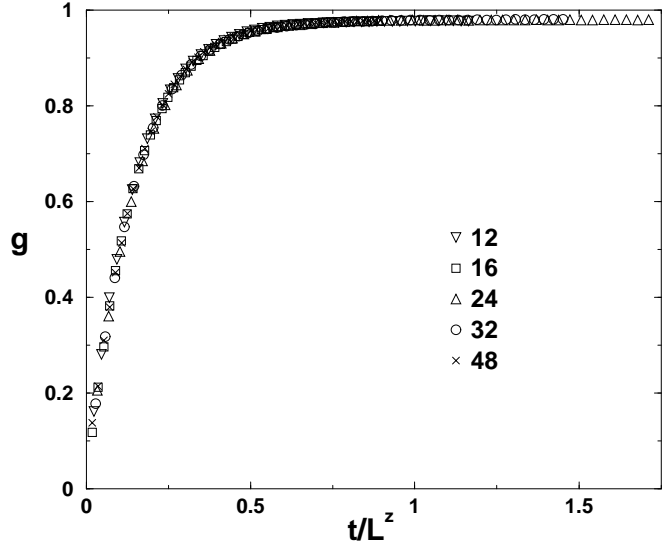}}
\caption{Scaling plot, with $z=2.35$, for the time-dependent Binder 
parameter, starting from a disordered initial condition, for system sizes 
$L=12,16,24,32,48$}. 
\label{xypow}
\end{figure}

At first sight, these results seem remarkable: Different values of $z$ 
are required to fit the approach to equilibrium from ordered 
(or `low-temperature') and disordered (or `high-temperature') initial 
states, whereas dynamical scaling predicts a unique value of $z$, namely 
that which describes equilibrium correlations (in this case $z=2$). 
What is going on here? It is worth noting that for Ising systems, the  
two different initial conditions give compatible results \cite{Ising}. 
The data for the XY model in $d=2$ seem to point clearly to a breakdown 
of dynamical scaling. This is indeed our conclusion, but the breakdown 
is weaker than the naive fit shown in Figure 2 suggests. We will argue 
that, for a disordered initial condition, the characteristic length 
scale $\xi(t)$ 
grows as $(t/\ln t)^{1/2}$ rather than $t^{1/z}$. Before presenting the  
arguments,  we test this prediction in Figure 3, where $t/L^2\ln(t/t_0)$ 
is used as abscissa. The fit is excellent. The value $t_0=0.5$ was used  
for the short-time cut-off, but the fit is not too sensitive to this 
value. 

The quality of the scaling collapses in Figures 2 and 3 are comparable, 
but the fit used in Figure 3 has a theoretical underpinning. First, however, 
we note that the scaling form (\ref{scaling}), with $z=2$, follows from 
the spin-wave theory for an ordered initial state: No free vortices 
are present at $t=0$, and none gets generated by thermal noise for any 
$T \le T_{KT}$. The calculation of $C(r,t)$ is straightforward \cite{RB}, 
and gives precisely the scaling form (\ref{scaling}) with 
$\xi(t) = t^{1/2}$ and $f(x) = \exp[-\eta\,J(x)/2]$, where 
$J(x) = \int_0^{x^2/8} (dy/y)\,[1 - \exp(-y)]$ and $\eta=1/4$ for 
$T=T_{KT}$. 

\begin{figure}
\narrowtext
\centerline{\epsfxsize\columnwidth\epsfbox{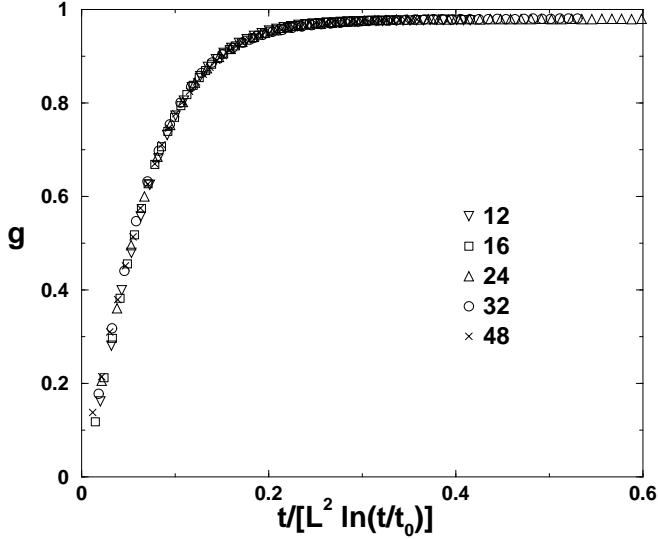}}
\caption{Same as Figure 2, but for scaling variable
$t/[L^2\ln(t/t_0)]$, with $t_0=0.5$.} 
\label{xylog}
\end{figure}

For a disordered initial condition, very different considerations are 
involved. The initial state contains many free vortices and antivortices. 
The approach to the equilibrium critical state proceeds through the 
annihilation of vortex-antivortex pairs, which is a slower process than 
the equilibration of spin waves. For pedagogical purposes, we consider 
first the case where the system evolves at $T=0$, instead of $T_{KT}$.
The evolution of the system via vortex-antivortex annihilation is an 
example of phase-ordering dynamics \cite{Review}. It is convenient to 
adopt a continuum approach based on the non-linear sigma model  
Hamiltonian $H = (1/2) \int d^2r\,(\nabla\vec{\phi})^2$, with local 
constraint $\vec{\phi}^2=1$. A field configuration describing a single 
free vortex, $\vec{\phi} = \vec{r}/|\vec{r}|$, has an energy 
$E_v = \pi\ln(L/a)$, where $L$ and $a$ are the system size and 
microscopic cut-off as before. A vortex-antivortex pair, separated 
by distance $R$, screen each other's far fields at scales larger than 
$R$, leading to a pair energy $E_p \simeq 2\pi\ln(R/a)$, and an 
attractive force $F = -dE_p/dR = -2\pi/R$ between the vortex and 
the antivortex. 

To discuss pair annihilation, some dynamics has to be imposed. 
The Monte-Carlo dynamics used here is in the `non-conserved' 
universality class (i.e.\ the magnetization is not conserved) 
described (at $T=0$) by the continuum model $\partial\vec{\phi}/\partial t 
= -\delta H/\delta\vec{\phi}$. This equation can be used \cite{Yurke} to 
compute an effective friction constant $\eta(R)$ associated with the 
motion of the vortex and antivortex under the force $F$ . 
An isolated vortex moving at speed $v$ in the $x$-direction has field 
configuration $\vec{\phi}(x,y,t) = \vec{\phi}_v(x-vt,y)$. Energy is 
dissipated at a rate $dE/dt = \int d^2r\,(\delta E/\delta \vec{\phi})\cdot
(\partial\vec{\phi}/\partial t) = 
-\int d^2r\,(\partial\vec{\phi}/\partial t)^2 =
-v^2 \int d^2r\, (\partial\vec{\phi}_v/\partial x)^2 = -\eta_v v^2$.  
Inserting the equilibrium vortex configuration, which is isotropic, gives  
the limiting zero-velocity friction constant as $\eta_0 = E_v$, i.e.\ 
$\eta_0$, like the vortex energy $E_v$, diverges logarithmically with the 
system size, $L$. For a vortex-antivortex pair, this translates into a 
logarithmic dependence on the separation \cite{Yurke}, $\eta(R) = 
\pi \ln (R/a)$. 

In the many-vortex situation envisaged for the nonequilibrium critical 
dynamics, the usual scaling arguments \cite{Review,Yurke,RB2} can be invoked, 
in which the pair separation, $R$, is replaced by the typical spacing, 
$\xi(t)$, between vortices and antivortices. The typical force on a vortex 
(or antivortex) is then $F \sim 1/\xi$, while the typical friction 
constant is $\eta \sim \ln(\xi/a)$. so the typical speed of a vortex 
is $d\xi/dt \sim F/\eta \sim 1/[\xi\ln(\xi/a)]$, giving 
$\xi(t) \sim [t/\ln (t/t_0)]^{1/2}$, with $t_0 \sim a^2$. An alternative 
approach leading to the same result is given in \cite{RB2}. 

For all $T$ in the range $0 \le T \le T_{KT}$, the large-scale properties 
in equilibrium are controlled by a fixed point with zero vortex fugacity, 
i.e.\ by the spin-wave theory, where the role of bound vortex-antivortex 
pairs is to renormalize the spin-wave stiffness. In the nonequilibrium 
case where free vortices and antivortices are present, due to a 
disordered initial condition, the dynamics on scales less than $\xi(t)$ 
should therefore be described by renormalized spin-wave theory, and the 
result $\xi(t) \sim [t/\ln (t/t_0)]^{1/2}$ should apply to {\em all} 
temperatures $T \le T_{KT}$, including $T_{KT}$ itself. 
This is our interpretation of the data in Figure 3. It accounts for 
the good data collapse using the appropriate scaling variable. 

Clearly our result, $\xi(t) \sim (t/\ln t)^{1/2}$ for a 
disordered initial condition, is asymptotically equivalent to an exponent 
$z=2$ (though the logarithmic correction still represents a scaling 
violation). So as $L$ and $t$ are increased we would expect the effective 
exponent, obtained by forcing a fit with a scaling variable $t/L^z$, to 
decrease towards 2. Collapsing data for pairs of $L$ gives effective 
exponents $z(L_1,L_2)$ given by $z(12,16) = 2.47(3)$, $z(16,24)=2.37(3)$, 
$z(24,32) = 2.29(3)$, and $z(32,48) = 2.34(3)$. The quoted errors are 
subjective. They are estimated from the quality of the data collapse, but 
make no allowance for statistical errors in the data. They therefore 
represent lower bounds on the true errors \cite{Note}. With this 
caveat the overall decreasing trend of the effective $z$ with 
increasing $L$ is clear, and accords with our expectations. 

Although the data presented here are restricted to the Kosterlitz-Thouless 
transition temperature, $T_{KT}$, the theoretical interpretation we have 
outlined holds for all $T \le T_{KT}$. In reference \cite{Zheng2}, data 
were obtained for a range of temperatures at and below $T_{KT}: T=0.90, 0.86, 
0.80$, and $0.70$. For a uniform initial state, the corresponding values 
of $z$ are $1.96(4), 1.98(4), 1.94(2)$, and $1.98(4)$, consistent 
with the result $z=2$, for all $T\le T_{KT}$, expected from spin-wave theory. 
The equivalent effective exponents obtained with a disordered initial state 
are consistently larger:  $z=2.29(1), 2.31(2), 2.33(1)$, and $2.38(2)$. 
We have argued that the correct interpretation of these anomalously large 
$z$ values is a logarithmically modified growth, 
$\xi(t) \sim [t/\ln(t/t_0)]^{1/2}$, of the characteristic length scale. 
The slow increase of the effective exponent with decreasing $T$ can be   
accounted for by a weak temperature dependence of the time-scale $t_0$ 
inside the logarithm.   

For the ordered initial condition, the scaling function $G(x)$ in 
equation (4) can, in principle, be calculated exactly using the 
spin-wave theory. This is technically more difficult, however, than 
the calculation \cite{RB} of the pair correlation function, because the 
evaluation of $\langle (\vec{M}^2)^2 \rangle$ involves 4-point 
correlation functions. We hope to present a detailed theory for $G(x)$ 
in future work. 

To summarize, we have argued that the rate of approach to equilibrium at 
(and below) the Kosterlitz-Thouless transition temperature depends on 
whether or not the initial state contains unbound vortices. Thus for a 
disordered initial state, where free vortices are present, the relaxation 
to equilibrium is slower, by logarithmic factors, than for an ordered 
initial state where no free vortices are present. It is possible that 
this result is peculiar to systems with defect-driven phase transitions. 
It goes against the expectation \cite{Janssen} that the scale length 
$\xi(t)$ controlling the relaxation to equilibrium, e.g.\ in (1) and (4), 
should be independent of the initial conditions (although the corresponding 
scaling {\em functions} may not be). This expectation is based on a 
perturbative renormalization group treatment in $4-\epsilon$ dimensions. 
Such an approach is not, presumably, sensitive to the effects of 
topological defects (vortices, in this case), which are the source of 
the scaling violations reported here.

A. Bray thanks M. A. Moore for a helpful discussion.

\end{multicols}

\end{document}